\documentclass[prb,preprint,superscriptaddress,groupedaddress]{revtex4}
\usepackage{graphicx}
\usepackage[caption=false,labelfont=bf]{subfig}
\usepackage{dcolumn}
\usepackage{bm}
\usepackage{comment}
\usepackage{color}
\usepackage{amsmath}
\usepackage{accents}
\usepackage[utf8]{inputenc}
\usepackage{bm}
\usepackage{float}

\newcommand{\be}{\begin{equation}}
\newcommand{\ee}{\end{equation}}
\newcommand{\bea}{\begin{eqnarray}}
\newcommand{\eea}{\end{eqnarray}}
\newcommand{\atan}{\mathrm{atan}}

\begin{document}
\title{Reliable spin-transfer torque driven precessional magnetization reversal with an adiabatically decaying pulse}
\author{D.~Pinna}
\email{daniele.pinna@nyu.edu}
\affiliation{Unit\'e Mixte CNRS/Thales, 91767 Palaiseau, France}
\affiliation{Department of Physics, New York University, New York, NY 10003, USA}
\author{C.~Ryan}
\affiliation{Raytheon BBN Technologies, Cambridge, MA  02138, USA}
\author{T.~Ohki}
\affiliation{Raytheon BBN Technologies, Cambridge, MA  02138, USA}
\author{A.~D.~Kent}
\affiliation{Department of Physics, New York University, New York, NY 10003, USA}
\date{\today}

\begin{abstract}
We show that a slowly decaying current pulse can lead to nearly deterministic precessional switching in the presence of noise. We consider a biaxial macrospin, with an easy axis in the  plane and a hard axis out-of-the plane, typical of thin film nanomagnets patterned into asymmetric shapes. Out-of-plane precessional magnetization orbits are excited with a current pulse with a component of spin polarization normal to the film plane. By numerically integrating the stochastic Landau-Lifshitz-Gilbert-Slonczewski equation we show that thermal noise leads to strong dephasing of the magnetization orbits. However, an adiabatically decreasing pulse amplitude overwhelmingly leads to magnetization reversal, with a final state that {\em only} depends on the pulse polarity, not on the pulse amplitude. We develop an analytic model to explain this phenomena and to determine the pulse decay time necessary for adiabatic magnetization relaxation and thus precessional magnetization switching.
\end{abstract}

\pacs{Valid PACS appear here}
\maketitle

\section{Introduction}

The study of spin current driven magnetic excitations has been a very active area of research over the past decade and has significant technological applications~\cite{Ralph2008,BrataasKentOhno2012}.  Specifically, the excitation of magnetization precession has led to current controlled oscillators that operate at 
GHz frequencies~\cite{Silva2008} and spin-current driven magnetization reversal has led to the development of non-volatile magnetic memory 
devices~\cite{KentWorledge2015}. Spin currents create spin-transfer torques (STT) that provide a means of exciting and driving non-linear magnetization dynamics
of nanometer scale magnets (or nanomagnets). The magnetization dynamics is also strongly affected by the presence of thermal noise which can alter the stability of 
magnetization states and the nature of the spin-transfer induced dynamics, including precessional magnetization orbits. 

Typically the magnetization dynamics consist of a fast gyromagnetic precession, whose amplitude
slowly changes over time due to spin torque and thermal effects. This separation of timescales can be used to study analytically the dynamical and thermal stability of nanomagnets subject to spin-polarized currents~\cite{Newhall2013,Pinna2013,Pinna2014}, and allows 
for a reduction in complexity of the stochastic Landau-Lifshitz-Gilbert-Slonczewski (sLLGS) equations to a simpler
one-dimensional stochastic differential equation. In the absence of damping, spin torque, and thermal noise, the dynamics conserve the macrospin energy, but in their presence a macrospin's dynamical evolution deviates from a constant energy trajectory.  

Thus an analysis of the noise-induced dynamics obtained by averaging the magnetization equations over constant-energy orbits provides significant new insights. Some of the authors of this paper have done this for a biaxial nanomagnet with an easy axis in the film plane and a hard axis out of the plane, typical of thin film nanomagnets patterned into asymmetric shapes (e.g. an ellipse)~\cite{Pinna2013,Pinna2014}.  Relevant dynamical scenarios have been shown to depend on the ratio between hard and easy axis anisotropies. The range of currents for which limit cycles exist was found, and the constant energy orbit averaging approach was used to study the magnetization dynamics of spin-torque oscillators, both in the
presence of thermal noise and as a function of the spin-polarization angle in a biaxial macrospin model~\cite{Pinna2013}. For this case analytical expressions were derived for currents that generate and sustain the out-of-plane precessional states. Further, there is a critical angle of the spin polarization necessary for the occurrence of such states. We also predicted a hysteretic response to applied current~\cite{Pinna2014}, which were tested in experiments on orthogonal spin-transfer devices~ \cite{Kent2004}, where the predicted hysteretic transitions into an intermediate resistance state were observed~\cite{Ye2015}.

Here we consider STT magnetization switching that occurs by out-of-plane precessional magnetization dynamics, as can occur in an orthogonal spin-transfer device. For this case it was widely thought that thermal and other noise sources would lead to dephasing of the precessional motion and thus an indeterminate magnetic state after the pulse ends. Here we demonstrate and explain a rather unexpected result that when the decay of a spin-current pulse is sufficiently slow the switching is very reliable even in the presence of noise, with the current pulse polarity determining the final magnetization state. After introducing our model for a biaxial nanomagnet and its dynamical modes we consider the effect of the pulse decay on the magnetization's final state. We determine the switching probability by numerically integrating the sLLGS equations and then describe our analytic model which explains the origin of the highly reliable switching. 

\section{Macrospin Model with Spin-Transfer Torques}

We study a macrospin with magnetization $\mathbf{M}$ of constant modulus ($M_s=|\mathbf{M}|$) with a biaxial magnetic anisotropy, with easy direction along the $\mathbf{\hat{x}}$-axis and hard direction along $\mathbf{\hat{z}}$. Its energy landscape depends on the projection of the magnetization onto these two axes\footnote{The projection along a third orthogonal axis is a dependent quantity arising from our choice of a fixed modulus magnetization.}. We write the easy and hard axis anisotropy energies as $K_E=\mu_0 M_sH_KV/2$ and $K_H=\mu_0 M_\mathrm{eff}^2V$, where $H_K$ is the anisotropy field, $M_\mathrm{eff}$ is the effective easy-plane anisotropy field, which is of order $M_s$ when this anisotropy has its origin only in the shape of the magnetic element~\cite{Chaves2015}, and $V$ is the volume of the magnetic element. In the absence of external magnetic fields and magnetic dipole fields arising from other magnetic layers, the energy can be written as:
\be
\label{eq:E_land}
E(\mathbf{m})=K_E\left[Dm_z^2-m_x^2 \right],
\ee
where $\mathbf{m}=\mathbf{M}/|\mathbf{M}|$ is the normalized magnetization vector and $D\equiv K_H/K_E=M_\mathrm{eff}^2/(M_sH_K)$ is a dimensionless ratio of the hard and easy axis anisotropies. $m_x$ and $m_z$ are the projections of the normalized magnetization vector on the x and z axis, i.e. $\mathbf{m}\cdot\mathbf{\hat{x}}$ and $\mathbf{m}\cdot\mathbf{\hat{z}}$ respectively. This energy has minima and thus stable magnetic configurations for $\mathbf{m}$ parallel and antiparallel to $\mathbf{\hat{x}}$, with an energy barrier $U=K_E$ separating these states. The out-of-equilibrium dynamics are described by the sLLGS equation:
\begin{equation}
\label{eq:Langevindynamics}
\dot{m}_i=A_i(\mathbf{m})+B_{ik}(\mathbf{m})\circ H_{th,k}
\end{equation}
where the stochastic contribution $\mathbf{H}_{th}$ is taken to have zero mean and delta-function correlation $\langle H_{th,i}(t)H_{th,k}(t')\rangle=2C\delta_{i,k}\delta(t-t')$. The diffusion constant $C=\alpha/(2(1+\alpha^2)\xi)$, with $\xi\equiv U/k_BT$ the energy barrier height divided by the thermal energy, is chosen to satisfy the fluctuation-dissipation theorem, and multiplicative noise `$\circ H_{th,k}$' is interpreted in the Stratonovich sense~\cite{Karatsas}. The expressions for the drift vector $\mathbf{A}(\mathbf{m})$ and diffusion matrix $\hat{\mathbf{B}}(\mathbf{m})$ terms read:
\begin{eqnarray}
\mathbf{A}(\mathbf{m})&=&\mathbf{m}\times\mathbf{h}_{\mathrm{eff}}-\alpha\mathbf{m}\times\left(\mathbf{m}\times\mathbf{h}_{\mathrm{eff}}\right)\nonumber\\
&-&\alpha I\mathbf{m}\times\left(\mathbf{m}\times\mathbf{\hat{n}}_p\right)-\alpha^2 I\mathbf{m}\times\mathbf{\hat{n}}_p,\nonumber\\
B_{ik}(\mathbf{m})&=&\sqrt{\frac{\alpha}{2\xi(1+\alpha^2)}}[-\epsilon_{ijk}m_j-\alpha(m_i m_k - \delta_{ik})],
\end{eqnarray}
where $\mathbf{h}_{\mathrm{eff}}=-\frac{1}{\mu_0 M_s H_K V}\nabla_{\mathbf{m}}E(\mathbf{m})$ is the effective field rescaled by $H_K$ and $\alpha$ is the Landau damping constant. STT effects due to current density $J$ (in units of A/m$^2$) are written in terms of a rescaled dimensionless current $I=(\hbar/2e)\eta J/(\alpha \mu_0 M_s H_K t)$, where $t$ is the thickness of the magnetic free layer and $\eta = (J_{\uparrow}-J_{\downarrow})/(J_{\uparrow}+J_{\downarrow})$ is the spin polarization along $\mathbf{\hat{n}}_p$. The effect of a STT depends on $\omega$, the angle between the spin-polarization axis $\mathbf{\hat{n}}_p$ and the easy axis $\mathbf{\hat{x}}$~(see Fig.~\ref{F1}(a))\footnote{A tilted spin-polarization axis allows modeling a spin-torque that results from more than one ``polarizing'' layer in a spin-valve (or MTJ) stack or, more generally, a free layer that has an easy-axis tilted relative to the spin-polarization axis.}. The temporal derivatives appearing in~(\ref{eq:Langevindynamics}) and throughout this paper are with respect to the natural timescale $\tau=\gamma \mu_0 H_K t/(1+\alpha^2)$, where $\gamma$ is the gyromagnetic ratio. The dynamics associated with Eqn.~\ref{eq:Langevindynamics} \cite{Palacios,LiZhang,Apalkov} leads to a Boltzmann equilibrium distribution of the magnetization at long times. 

Under the assumption that the precessional timescale is much smaller than that of damping, spin-transfer torque and thermal diffusion, Eqn.~\ref{eq:Langevindynamics} can be effectively reduced to a 1D stochastic differential equation for the evolution of the macrospin's instantaneous energy $E$ as a function of time~\cite{Newhall2013,PinnaIEEE}. This has proven useful because it allows the macrospin's dynamics to be characterized analytically in many interesting physical situations, which we now summarize.

\subsection{Biaxial Macrospin Model}
\begin{figure}
         \begin{centering}
    	\subfloat[]{{\includegraphics[width=2.0in]{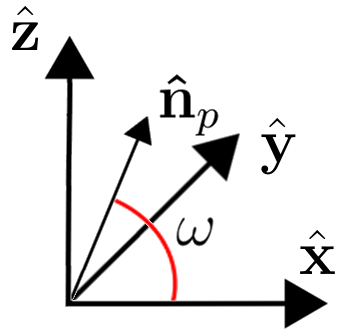} }}%
    	\qquad
    	\subfloat[]{{\includegraphics[width=3.4in]{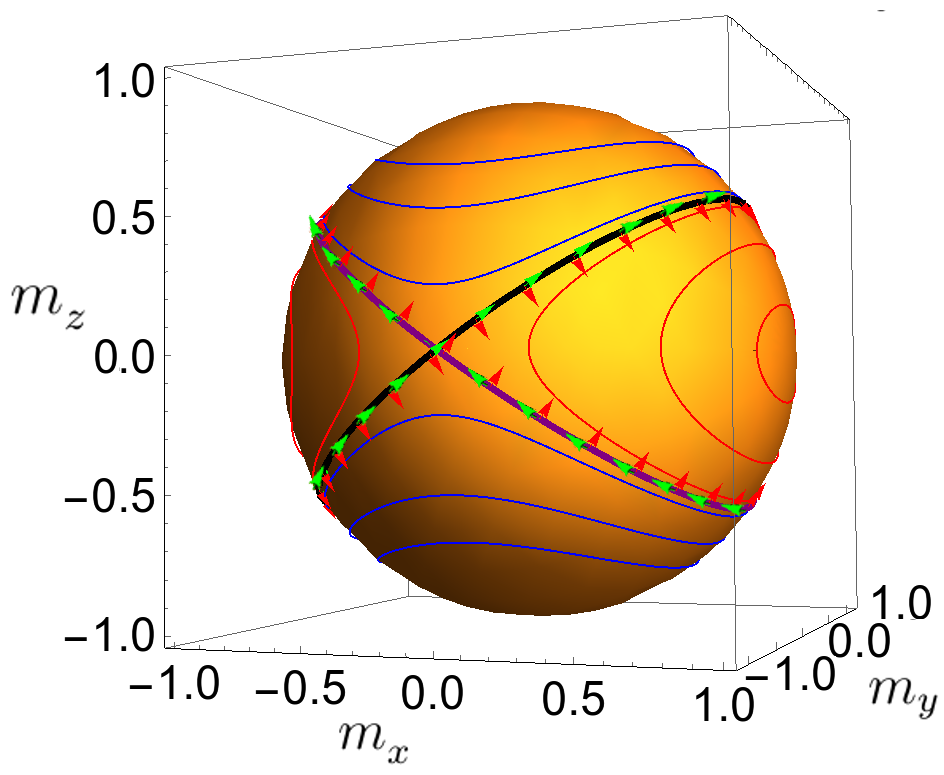} }}%
 \end{centering}
\caption{{\footnotesize(a) Uniaxial easy ${\bf \hat{x}}$ and hard-axis ${\bf \hat{z}}$ magnetic anisotropy directions are shown along with spin-polarization direction ${\bf \hat{n}_p}$. The spin-polarization makes an angle $\omega$ to the easy axis. (b) Constant energy trajectories for $D=10$. $\epsilon<0$ trajectories are shown in red whereas $\epsilon>0$ trajectories are shown in blue. Notice how two distinct basins exist for positive and negative energy trajectories. The separatrix, corresponding to $\epsilon=0$, can be parametrized as two intersecting circles $\bm{\gamma}^{1,2}(s)$ shown in black and purple respectively. Their tangents $\bm{\gamma}_{//}^{1,2}(s)$ and normal components $\bm{\gamma}_{\bot}^{1,2}(s)$ are indicated by green and red arrows respectively.}}

\label{F1}
\end{figure}

In the absence of damping, spin-torque and thermal noise, the dynamics~(\ref{eq:Langevindynamics})
preserve the macrospin's energy which, expressed in dimensionless form, reads:	
\be
\label{eq:epsilon1}
\epsilon=\frac{E(\mathbf{m})}{U}=Dm_z^2-m_x^2.
\ee
The conservative trajectories come in two different types. For $-1<\epsilon<0$ the magnetization gyrates
around the easy $\mathbf{\hat{x}}$-axis and is said to be precessing ``in-plane" (IP). For $0<\epsilon<D$, the magnetization precesses about the hard $\mathbf{\hat{z}}$-axis and is said to be precessing ``out-of-plane" (OOP). A sample of these trajectories for positive and negative energies is shown in Fig.~\ref{F1}(b). The dashed black line represents the separatrix which divides the Bloch sphere into four distinct dynamical basins: two $\epsilon<0$ IP basins, and two $\epsilon>0$ OOP basins. 

\begin{figure}
	\begin{center}
	\centerline{\includegraphics [clip,width=4in, angle=0]{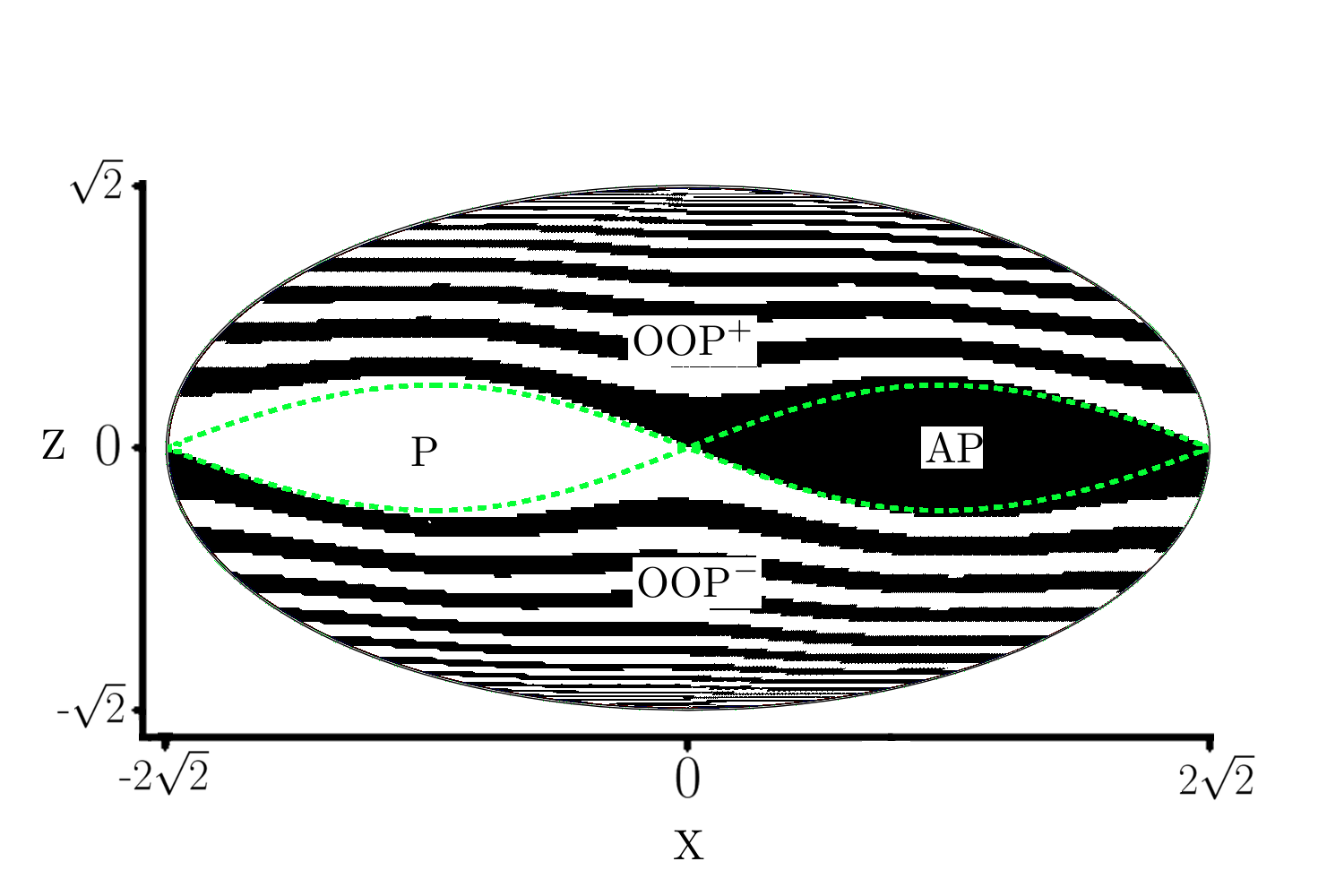}}
	\end{center}
	\caption{{\footnotesize Bloch sphere representation of the zero temperature relaxed configuration as a function of initial magnetization for $D=10$ and $\alpha=0.04$. $X$ and $Y$ correspond to coordinates of the Bloch sphere in a Molleweide spherical projection~\cite{Molleweide}. White and black correspond to P and AP final states respectively. The dashed green line shows the dynamical separatrix between the $\epsilon<0$ P ($m_x<0$) and AP ($m_x>0$) basins and $\epsilon>0$ $\mathrm{OOP}^+$ ($m_z>0$) /$\mathrm{OOP}^-$ ($m_z<0$) basins.}}
	\label{F2}
\end{figure}

\begin{figure}%
    \begin{centering}
    \subfloat[]{{\includegraphics[width=2.7in]{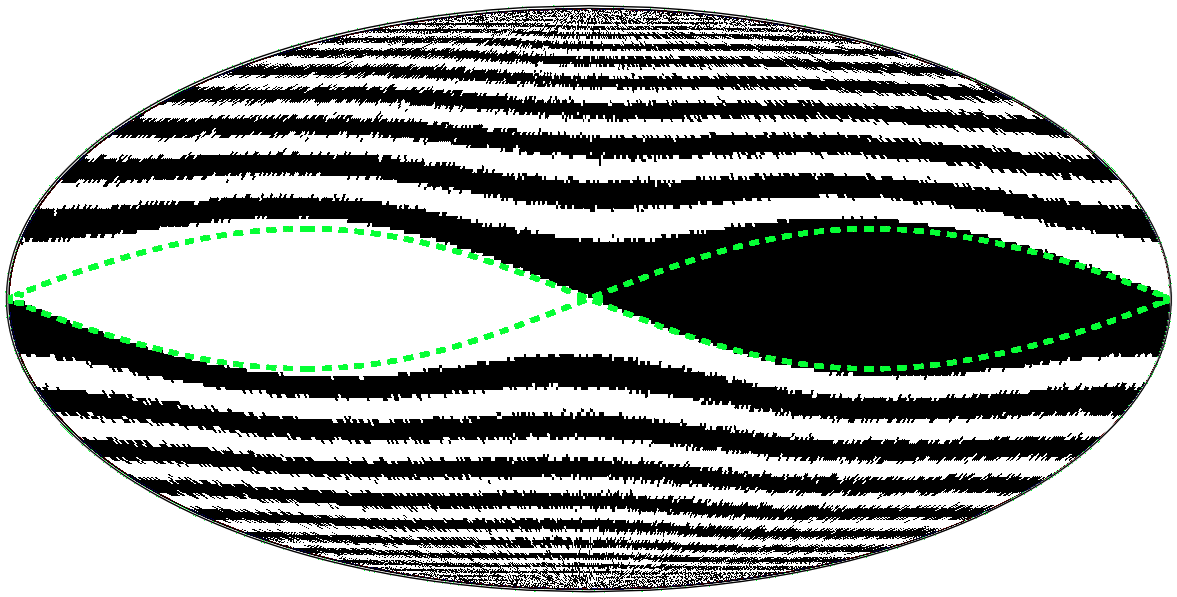} }}%
    \qquad
    \subfloat[]{{\includegraphics[width=2.7in]{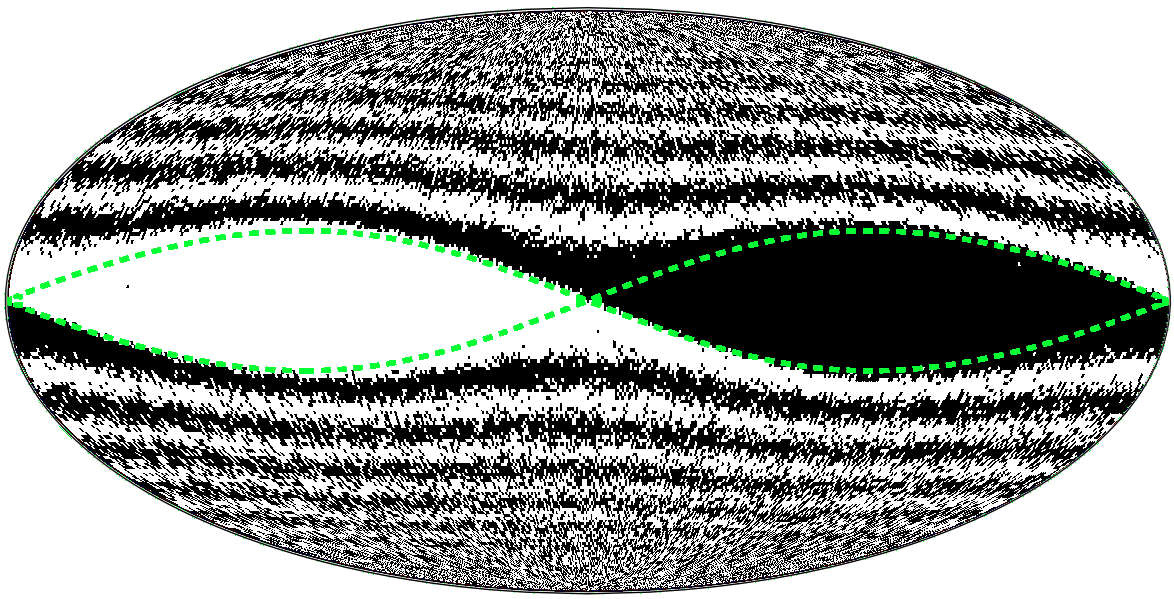} }}%
    \qquad
    \subfloat[]{{\includegraphics[width=2.7in]{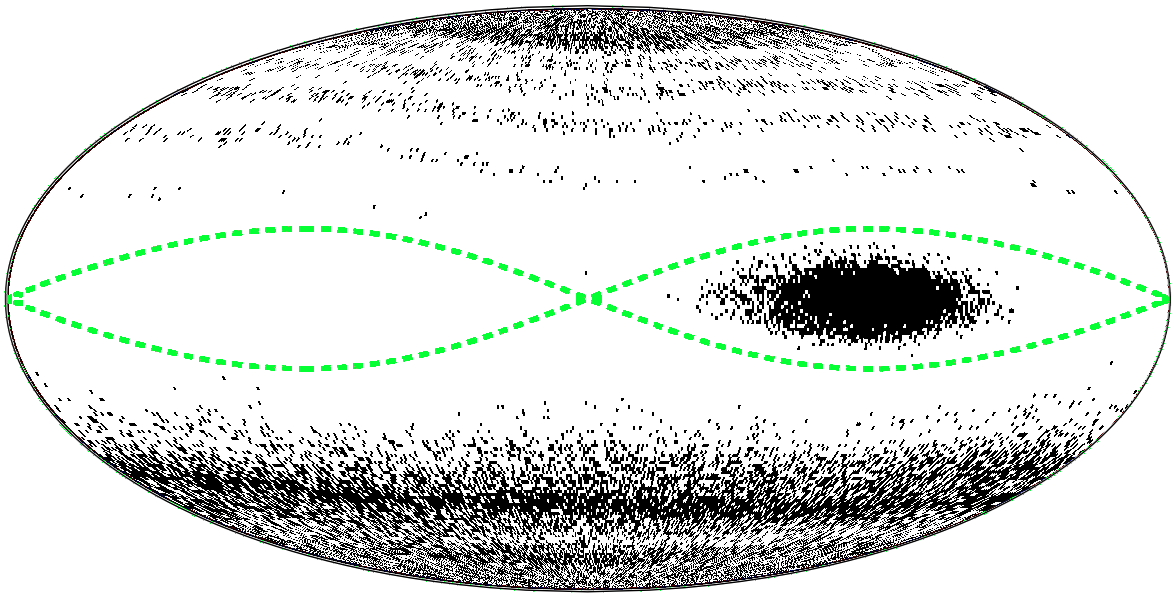} }}%
    \qquad
    \subfloat[]{{\includegraphics[width=2.7in]{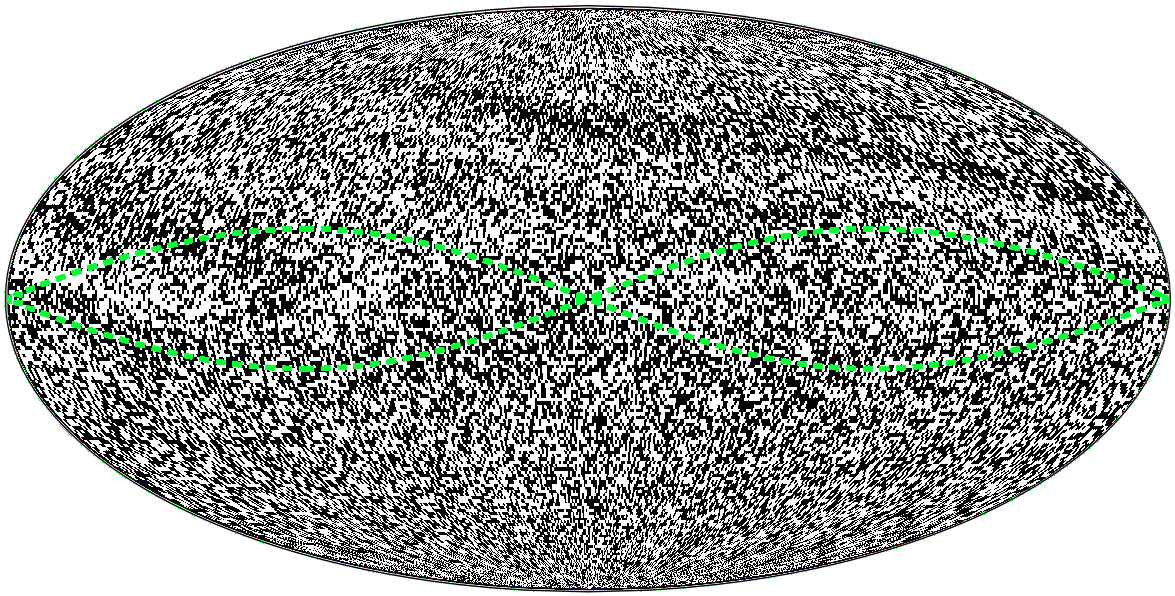} }}%
    \end{centering}
    \caption{{\footnotesize Relaxed configuration as a function of initial magnetization for $D=10$ and $\alpha=0.04$ where colors white/black correspond to relaxation into P/AP basins respectively. Both (a) and (b) correspond to a scenario where the macrospins are allowed to relax without applied current in the presence of thermal effects with $\xi=1200$ and $\xi=80$ respectively (larger $\xi$ implies lower temperature). Subfigures (c) and (d) correspond to relaxation at temperature $\xi=80$ preceded by a constant $1\;\mathrm{n}s\cdot T$ current pulse (physical times are obtained upon dividing by $\mu_0H_K$)
of intensity $I=1.5\,I_c$ and axial tilt $\omega=0.5\,\omega_c$ and $\omega=2.5\,\omega_c$ respectively.}}%
    \label{F3}%
\end{figure}

Upon introducing the effects of damping, the dynamics will dissipate magnetic energy, thus mapping any initial state of the configuration sphere into a corresponding final state either aligned parallel (P) or antiparallel (AP) with the easy $\mathbf{\hat{x}}$-axis. Figure~\ref{F2} shows a projectional map of the Bloch sphere color coded according to the state to which the magnetization relaxes; $D=10$ and $\alpha=0.04$ was chosen for the plot and white/black regions correspond to P/AP final states respectively. 
In Fig.~\ref{F3}(a) \& (b), the Bloch sphere of an identical macrospin model relaxes in the presence of thermal noise with intensity $\xi=1200$ and $\xi=80$ (larger $\xi$ corresponds to lower temperature). We omit the Molleweide axes labels in this and subsequent figures as they are identical to those used in Fig.~\ref{F2}.
Thermal effects can be seen to modify the zero temperature relaxation shown in Fig.~\ref{F2} by blurring the boundaries of the white and black regions; the relaxation process becomes stochastic.

The introduction of a driving current will strongly affect the magnetization dynamics due to the additional spin-transfer torque biasing either the P or AP basins. We note that this is generally a non-conservative torque and thus its effects cannot be described in terms of the gradient of an effective energy. This renders many techniques used to analyze the energetics involved in the macrospin's evolution inapplicable. However, previous work has shown that whenever the timescales for thermal and spin-torque driven diffusion are much larger than the conservative precessional timescale, non-conservative effects can be studied perturbatively~\cite{Apalkov,Newhall2013,Pinna2013}. Effectively, the macrospin precesses multiple times along nearly constant energy trajectories, only diffusing slowly in energy. This allows for an averaging of the LLGS dynamics (\ref{eq:Langevindynamics}) along constant energy trajectories (shown in Fig.~\ref{F1}(b)) to obtain a description of the macrospin's dynamics solely in terms of diffusive behavior over its conservative energy landscape~\cite{Pinna2013}. An analysis of the time evolution of the macrospin's energy provides significant insights into magnetization dynamics in the presence of STT.

We summarize the main results of such an analysis, the details of which can be found in Refs.~\cite{Pinna2013,Pinna2014}. There are two 
fundamental features of a biaxial macrospin subject to a spin current. The first is that there are two critical currents ($I_c$ and $I_\mathrm{OOP}$). For currents $I>I_c=(2/\pi)\sqrt{D(D+1)}/\cos\omega$ the entire AP basin becomes unstable ($\dot{\epsilon}>0$ for all $\epsilon<0$) and magnetic states within it will be driven into the $m_z>0$ OOP basin ($\mathrm{OOP}^+$). In turn, the magnetization will either then proceed to relax to the stable P basin or, if $I>I_\mathrm{OOP}=I_c/(\sqrt{D}\tan\omega)$, remain in the $\mathrm{OOP}^+$ basin evolving along a constant energy orbit, thus maintaining a steady-state OOP precession. 

The second fundamental feature is there is a critical tilt of the spin-polarization axis that determines the nature of the magnetization dynamics excited by the spin-transfer torque. At the critical tilt $\omega_c=\atan(1/\sqrt{D})$, $I_c=I_\mathrm{OOP}$ and for $\omega>\omega_c$, $I_c>I_\mathrm{OOP}$. For subcritical tilts $\omega<\omega_c$ the current can be increased such that $I_c<I<I_\mathrm{OOP}$. If the current is increased sufficiently slowly (so that the constant energy orbit approach applies), magnetizations in the AP state will evolve toward OOP states but immediately relax into the P basin, leading to a deterministic switch~\footnote{$I_c$ is denoted the critical switching current for this reason.}. Conversely, for supercritical tilts, all magnetic states excited into OOP states will remain there until the current is lowered back below $I_\mathrm{OOP}$. 

These magnetization switching characteristics can be seen by numerically integrating the sLLGS equation, Eq.~\ref{eq:Langevindynamics}. We have done this for an ensembles of 92160 independent macrospins, sampling the entire Bloch sphere homogeneously, with an integration time step of $0.01$ in natural time units, i.e. $\tau$. Results are shown in Fig.~\ref{F3}(c) \& (d), where we take a damping constant $\alpha=0.04$ and $D=10$.

Fig.~\ref{F3} shows the final magnetization state after a current pulse greater than the $I_c$ ($I=1.5I_c$) for subcritical tilts ($\omega=0.5\omega_c$ Fig.~\ref{F3}(c)) and supercritical tilts ($\omega=2.5\omega_c$ Fig.~\ref{F3}(d)). In both cases the thermal noise $\xi=80$ is present and the driving current is instantaneously switched off at the end the $1$ ns$\cdot$T pulse (physical times are obtained upon dividing by $\mu_0H_K$). For subcritcial tilts a large fraction of initial magnetization states have switched into the P basin. The not-switched states are around the AP state $m_x=1$ and the north and south poles of the Bloch sphere. Those near the AP state did not have time to switch during the finite duration of the current pulse; had the pulse been left on for a longer time there would been fewer not-switched states in this zone. In the supercritical case the large tilt allows the current to excite all IP states into OOP orbits where thermal noise and dynamical decoherence shuffle the trajectories enough that, once relaxation takes place in the absence of a current, the final states appears random.


Having seen the characteristic switching dynamics in the presence of noise for sub- and supercritical tilts, we now investigate a case in which the driving current is switched off gradually as opposed in the stepwise fashion we have considered up to this point. We again consider a pulse that is turned on for a time $0.27 \,(n\mathrm{s}\cdot T)$ but decays exponentially, $I(t)=I_0\exp\left(-t/\tau_I\right)$. The pulse thus decays from a value $I_0>I_c>I_\mathrm{OOP}$ with a time constant $\tau_I$. We again sample the entire Bloch sphere and determine to what state the magnetization relaxated in the presence of noise. 
  
\begin{figure}[H]%
	\begin{center}
	\centerline{\includegraphics [clip,width=5in, angle=0]{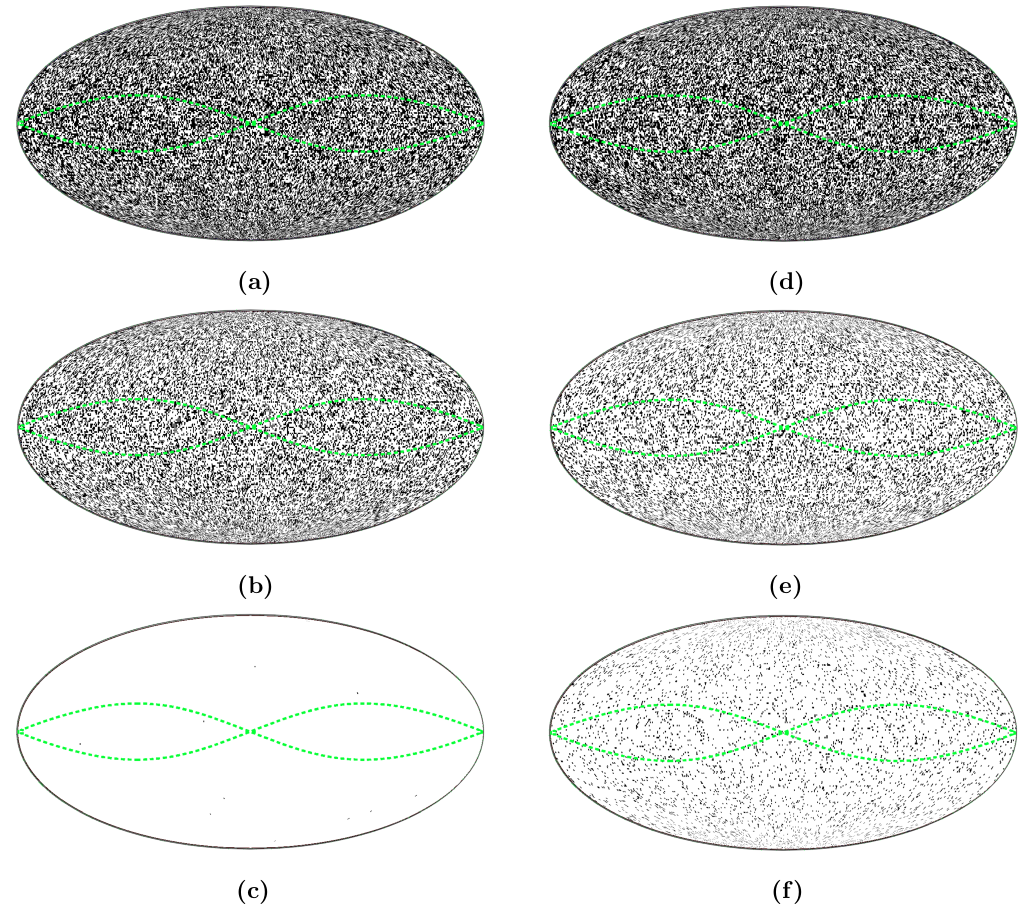}}
	\end{center}
    \caption{{\footnotesize Relaxed configuration of the Bloch sphere as a function of initial magnetization for $D=10$, $\alpha=0.04$ and $\omega=2.5\,\omega_c$ where colors white/black imply relaxation into P/AP basins respectively. We show the effect of an initially constant ($I_0=1.5\,I_c$) $0.27 \,(n\mathrm{s}\cdot T)$ current pulse followed by an exponential decay (with the pulse polarity favoring the P basin) for different temperatures $\xi=5714$ (left column, Figs:a-c) and $\xi=80$ (right column, Figs:d-f). From top to bottom, the exponential relaxation timescale $\tau_I=0.01,0.15,0.3\,(n\mathrm{s}\cdot T)$. (Physical times are obtained upon dividing by $\mu_0H_K$.) For slow enough current decays, the magnetization relaxes into the P basin nearly deterministically. The dashed green line shows the dynamical separatrix.}}
    \label{F4}%
\end{figure}
Figure~\ref{F4} shows the relaxation behavior for $D=10$, $\alpha=0.04$, $\omega=2.5\,\omega_c$ at different temperatures ($\xi=5714,80$ corresponding to left/right column respectively) and current decay timescales ($\tau_I=0.01,0.07,0.15\;(n\mathrm{s}\cdot T)$ from top to bottom). As the decay rate is made progressively larger, the Bloch sphere is seen to overwhelmingly relax into the P basin. This result is in remarkable contrast to what was seen in Figure~\ref{F3}(d). The current pulse is sufficient to excite OPP orbits yet if the pulse decay time is sufficiently slow the magnetization relaxes reliably into a state set by the current polarity (positive current in our model favors the P state). The effect is more pronounced at lower temperature (compare Fig.~~\ref{F4}(c) and (f)).

\begin{figure}
	\begin{centering}
        \includegraphics[width=3in]{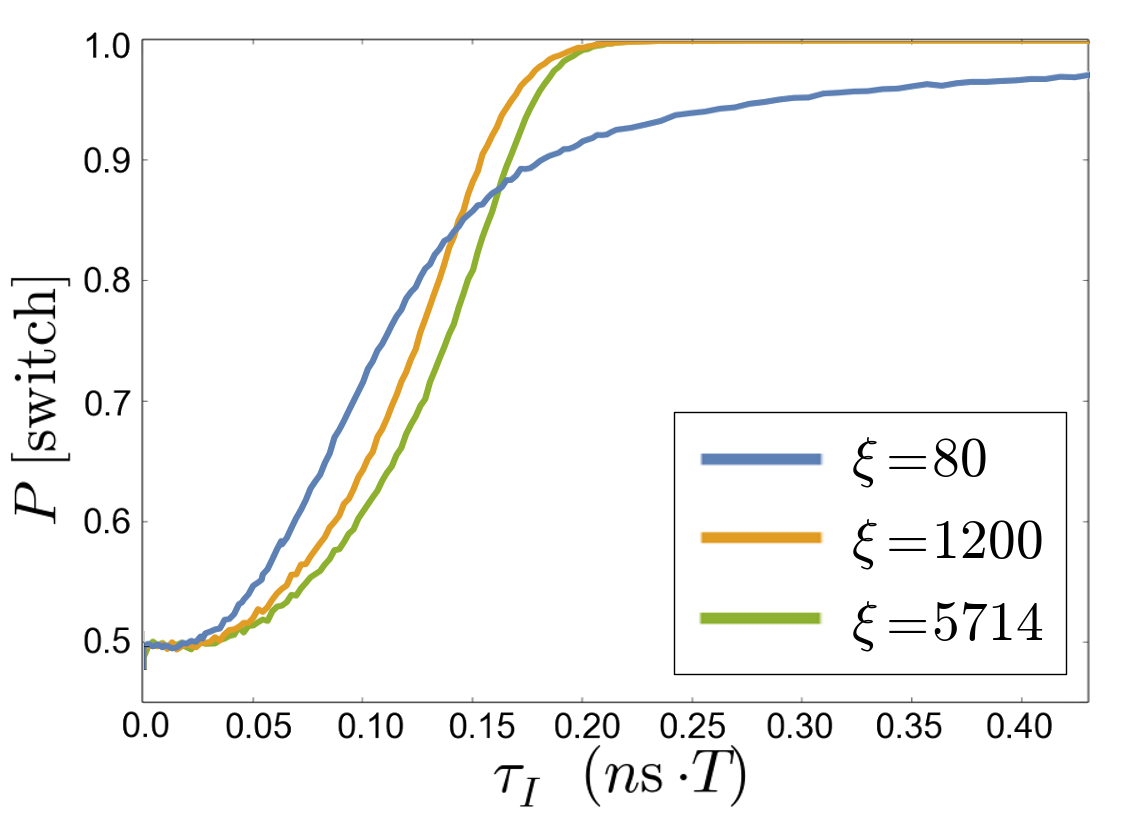}
    	\end{centering}
	\caption{{\footnotesize Switching probability as a function of pulse decay time for a model with $D=10$, $\alpha=0.04$ and $\omega=2.5\,\omega_c$. Physical time is obtained upon dividing by $\mu_0 H_K$. The switching probabilities were computed by first driving the magnetization with a constant ($I_0=1.5\,I_c$) $0.27 \,(n\mathrm{s}\cdot T)$ current pulse followed by an exponential decay. Larger temperatures lead to lower overall switching probability consistent with the increased thermal noise.}}
	\label{F5}
\end{figure}

To further highlight the role of the pulse relaxation time and the temperature in this phenomena, we show the switching probability for three different temperatures ($\xi=5714,1200$ and $80$) as a function of the pulse relaxation time $\tau_I$ in Fig.~\ref{F5}. It is clear that the effect is robust as a function of temperature but the switching probability increases on reducing the temperature. We further find that the switching probability is nearly independent of the pulse amplitude (i.e $I_0$) provided the pulse is of sufficient amplitude and duration to excite the vast majority of IP states into OPP orbits. This dynamics is thus an important and fundamental characteristic of a biaxial macrospin subject to a spin-transfer torque and demands a physical and mathematical explanation.

\section{Orthogonal Drift Biasing at the Separatrix}

To explain this phenomena we consider a scenario where a fixed current is sustaining a stable OOP precessional state ($I>I_\mathrm{OOP}$). As discussed in the previous section, the macrospin's steady-state dynamics trace a constant energy orbit, which is a fixed point of the averaged energy dynamics. Changing the current will alter the fixed point and the macrospin will diffuse to a new constant energy trajectory. If changes in the driving current are made slowly enough, then relaxational dynamics will ensure that the energy of the macrospin's precessional orbits evolves adiabatically with the current. In turn, the macrospin's average energy will trace a sequence of fixed points. When $I=I_\mathrm{OOP}$, the structure of the LLGS drift at the separatrix (averaged over one $\epsilon=0$ revolution) will then influence which IP basin ($m_x<0$ or $m_x>0$) the magnetization relaxes into.

To make these statements quantitative, we first parametrize the energy separatrix and project the complete LLGS dynamics~(\ref{eq:Langevindynamics}) onto both its tangent and normal. As shown in Fig.~\ref{F1}(b), the energy separatrix is an intersection of two great circles (shown in black and purple). We denote by $\bm{\gamma}^{1,2}(s)$ the circles composing the separatrix, where $s$ is the coordinate on the circle and $1,2$ correspond to the black/purple circle respectively. The tangent to the separatrix will be $\bm{\gamma}_{//}^{1,2}(s)=\partial_s\bm{\gamma}^{1,2}$ and its normal is given by $\bm{\gamma}_{\bot}^{1,2}(s)=\bm{\gamma}^{1,2}\times\bm{\gamma}_{//}^{1,2}$ (shown as green and red arrows respectively in Figure~\ref{F1}). By noting that at $\epsilon=0$ the magnetization components must satisfy $Dm^2_z=m^2_x$:
\bea
\label{eq:SepParam}
\bm{\gamma}^{1,2}(s)&=&\left(\pm\sqrt{\frac{D}{D+1}}\sin(s),\cos(s),\frac{1}{\sqrt{D+1}}\sin(s)\right)\\
\bm{\gamma}_{//}^{1,2}(s)&=&\left(\pm\sqrt{\frac{D}{D+1}}\cos(s),-\sin(s),\frac{1}{\sqrt{D+1}}\cos(s)\right)\\
\bm{\gamma}_{\bot}^{1,2}(s)&=&\frac{1}{\sqrt{D+1}}\left(1,0,\mp\sqrt{D}\right),
\eea 
with $s\in[0,2\pi]$ and $\bm{\gamma}^{1,2}(0)$ is a unit vector along the $y$-axis. Increasing $s$ traces the circle along  $\bm{\gamma}_{//}^{1,2}$.

Projecting the magnetization dynamics onto $\bm{\gamma}_{//}^{1,2}(s)$ and $\bm{\gamma}_{\bot}^{1,2}(s)$ gives (see Appendix A):
\bea
\label{eq:SepDynamics}
\dot{\mathbf{m}}\cdot\bm{\gamma}_{//}^{1,2}(s)&=&\mp2\sqrt{D}\left[\sin(s)+\frac{\alpha\sqrt{D}}{\pi}\frac{I}{I_c}\left(1\mp\tan\omega\tan\omega_c\right)\right]\\
\dot{\mathbf{m}}\cdot\bm{\gamma}_{\bot}^{1,2}(s)&=&2\alpha\sqrt{D}\left[\pm\sin(s)+\frac{1}{\pi}\frac{I}{I_c}\left(1\mp\frac{\tan\omega}{\tan\omega_c}\right)\right],
\eea
where the critical tilt, as mentioned earlier is, $\omega_c=\atan(1/\sqrt{D})$. (As a reminder, $\omega>\omega_c$ is assumed in this analysis, as this condition must be satisfied to have stable OPP precessional states.)

On the separatrix, where the period of conservative trajectories formally diverges~\cite{Pinna2014}, we see that the timescale for drifting across the separatrix is a factor of $1/\alpha$ ($\sim 20$) larger than that for precessing along it. We thus consider only the portion of the separatrix that bounds the $\mathrm{OOP}^+$ basin ($s\in[0,\pi]$) and compute the average net drift orthogonal to the separatrix:
\be
\label{eq:SeparatrixAverage}
\langle\dot{\mathbf{m}}\cdot\bm{\gamma}_{\bot}^{1,2}\rangle_{\mathrm{orbit}}=\frac{2\alpha\sqrt{D}}{\pi}\left[\pm 1+\frac{I}{I^C}\left(1\mp\frac{\tan\omega}{\tan\omega_C}\right)\right],
\ee
where $\langle\cdot\rangle_{\mathrm{orbit}}$ implies averaging over the coordinate range $s\in[0,\pi]$. Given the convention chosen for the orientation of the normals to the separatrix (see Fig.~\ref{F1}(b)), a positive average orthogonal flow ($\langle\dot{\mathbf{m}}\cdot\bm{\gamma}_{\bot}^{1,2}\rangle_{\mathrm{orbit}}>0$) will always bias exiting the separatrix into the P basin. This is the case whenever:
\be
\left(\frac{\tan\omega}{\tan\omega_c}+1\right)^{-1}<\frac{I}{I_c}<\left(\frac{\tan\omega}{\tan\omega_c}-1\right)^{-1}
\ee
Since $I_\mathrm{OOP}=I_c/(\tan\omega/\tan\omega_c)>I_c/(1+\tan\omega/\tan\omega_c)$, this biased orthogonal drift effect will {\em always} occur whenever the magnetization crosses the separatrix for $I=I_\mathrm{OOP}$.
We now proceed to show how this leads the magnetization to relax into a specific IP basin upon slowly reducing the current sustaining OOP precessional orbits. 

\section{Near Deterministic Relaxation}

\begin{figure}
	\begin{centering}
    	\includegraphics[width=3in]{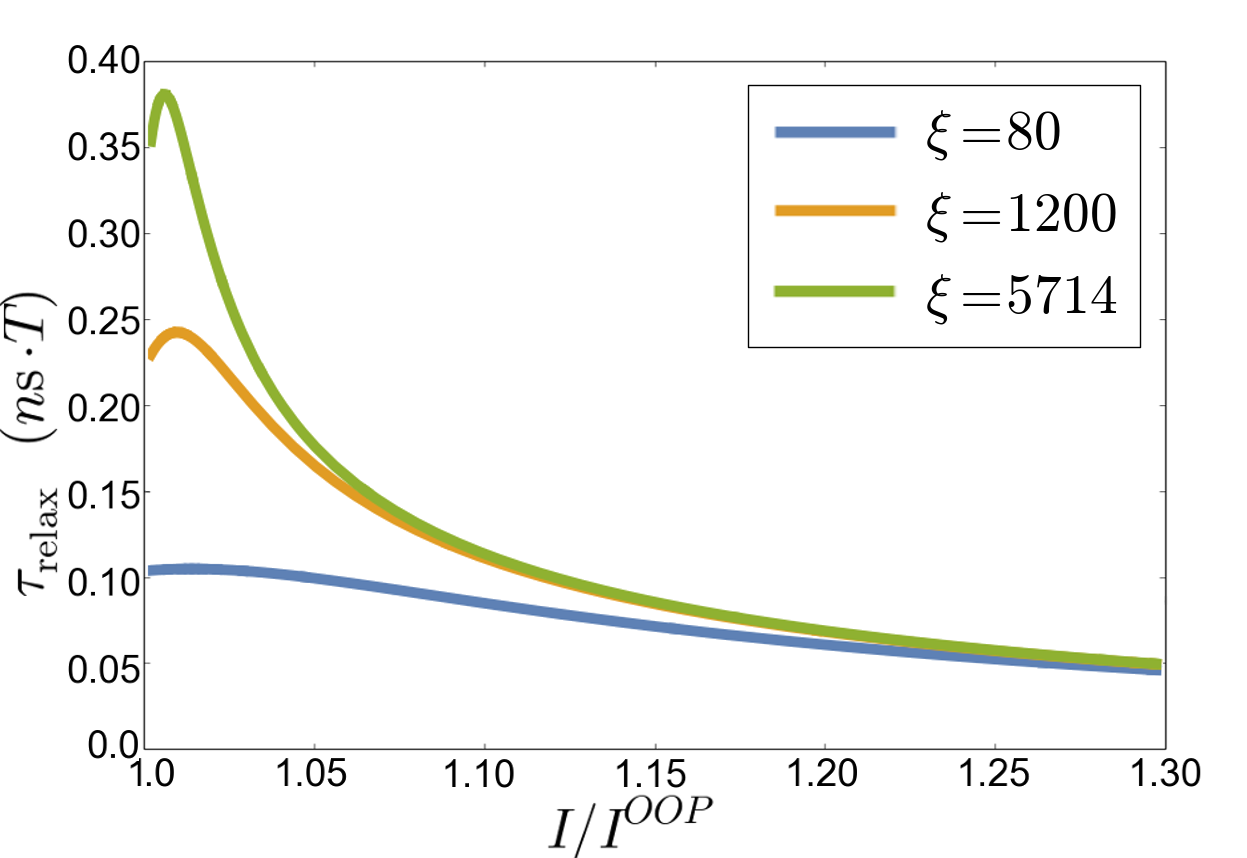}
        	\end{centering}
	\caption{{\footnotesize Predicted current and temperature dependence (larger $\xi$ implies lower temperature) of the energy relaxation rate for a model with $D=10$, $\alpha=0.04$ and $\omega=2.5\,\omega_c$. (Physical time is obtained upon dividing by $\mu_0 H_K$.) Whereas larger currents and temperatures ($\xi\propto 1/k_BT$) are found to favor faster relaxation of the magnetization dynamics to their steady-state equilibrium, larger temperatures lead to lower overall switching probability due to increased thermal noise present when crossing the separatrix, as shown in Fig.~\ref{F5}.}}
	\label{F6}
\end{figure}

We will now determine the timescales on which the magnetization will relax to its energy fixed point if perturbed to determine the requirements on the pulse decay time for reliable magnetization reversal. Upon linearizing the energy evolution equations around $\epsilon_0$ fixed point ($\epsilon\to\epsilon_0+\delta\epsilon$) (see Appendix B), the perturbations will be governed by dynamics which exponentially decay with timescale $\tau_{\mathrm{rel}}(I)$.
Figure~\ref{F6} shows how the relaxation rate  of perturbations to a given steady-state precessionary state is expected to change as a function of currents $I>I_\mathrm{OOP}$ and varying temperature for a sample with $D=10$, $\alpha=0.04$ and $\omega=2.5\,\omega_c$. Larger currents and temperature are seen to favor a faster relaxation of the magnetization dynamics.

If temporal variations of the current happen on a timescale $\tau_I>\tau_{\mathrm{rel}}$, the magnetization will quickly respond to any destabilizing effects and continuously trace nearly constant energy orbits (the adiabatic condition). If, on the other hand, $\tau_I<\tau_{\mathrm{rel}}$ the magnetization will be in an out-of-equilibrium state which cannot be characterized with energy averaging techniques. 

This model thus captures the physics of the situation we simulated numerically in Sec. II, a case in which the a current $I>I_\mathrm{OOP}$ initially sustains a stable OOP precessional orbit (Figs.~\ref{F4}\&\ref{F5}) and is reduced. If the current is decreased slowly enough for the adiabaticity conditions to be satisfied, the magnetization will experience an orthogonal drift biasing effect and nearly deterministically switch into the P or AP basin depending on the current's polarity. Noise perturbs this deterministic dynamics, causing the magnetization to occasionally jump into the unbiased IP basin even under adiabaticity conditions whenever the effective energy barrier separating the IP/OOP basins becomes comparable to the thermal noise strength. This will happen always as the magnetization precesses very close to the separatrix. For a more detailed quantitative exposition of these effects, the precise orbital behavior near the separatrix (and not just the constant energy orbit averaging description) must be taken into account, which is beyond the scope of this article.

At lower temperature, however, we see our model capturing the relevant time scales of the switching dynamics quite well. In fact, for current decay timescales comparable to the maximum relaxation timescale ($\tau_I\sim\mathrm{max}\{\tau_{\mathrm{relax}}\}$) we see that the switching probability plotted in Fig.~\ref{F5} approaches $1$ (e.g. for $\xi=5714$, $\mathrm{max}\{\tau_{\mathrm{relax}}\}\simeq 0.35\,n\mathrm{s}\cdot T$ and high switching probability is seen to take place for current decay timescales $\tau_I\geq 0.2\,n\mathrm{s}\cdot T$).

\section{Effect of Initial Current}
\begin{figure}[H]%
    \centering
    \subfloat[]{{\includegraphics[width=2.7in]{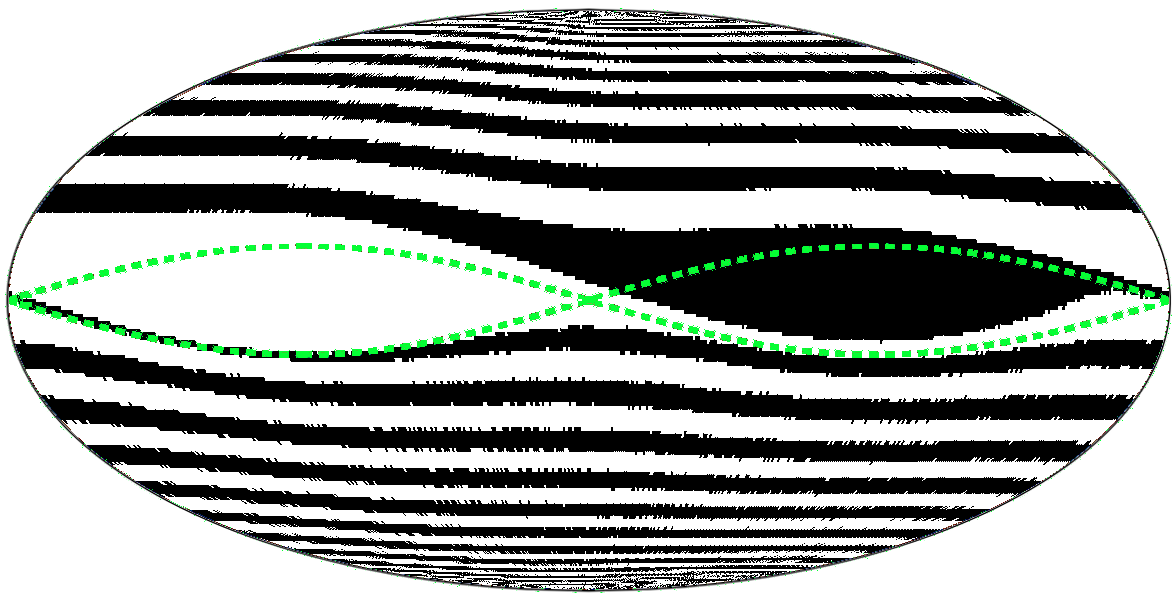} }}%
    \qquad
    \subfloat[]{{\includegraphics[width=2.7in]{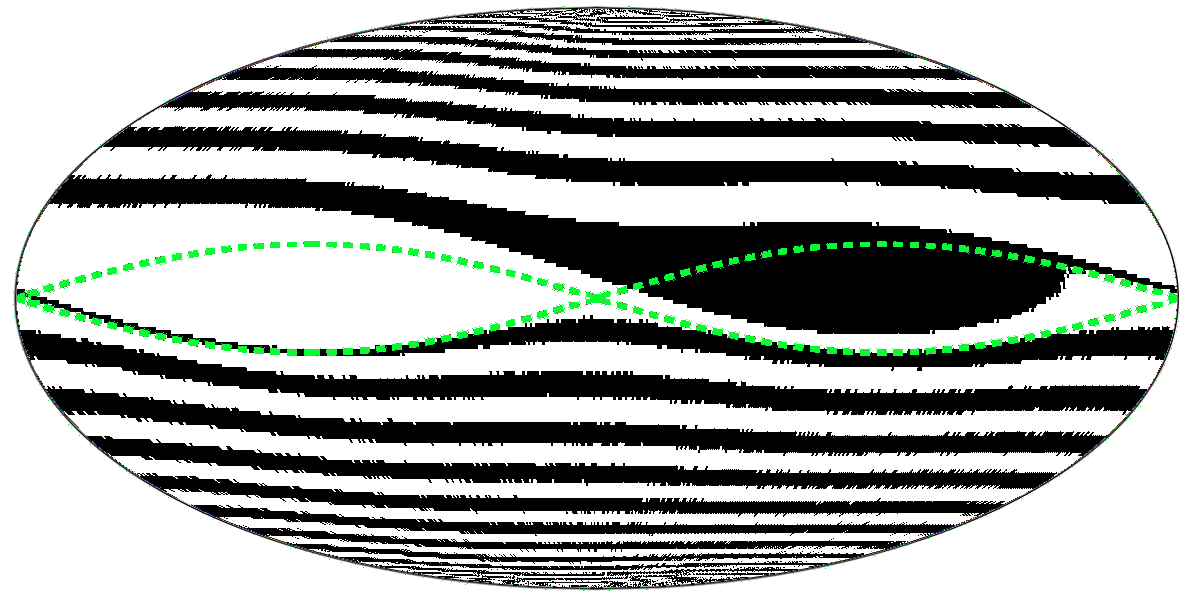} }}%
    \qquad
    \subfloat[]{{\includegraphics[width=2.7in]{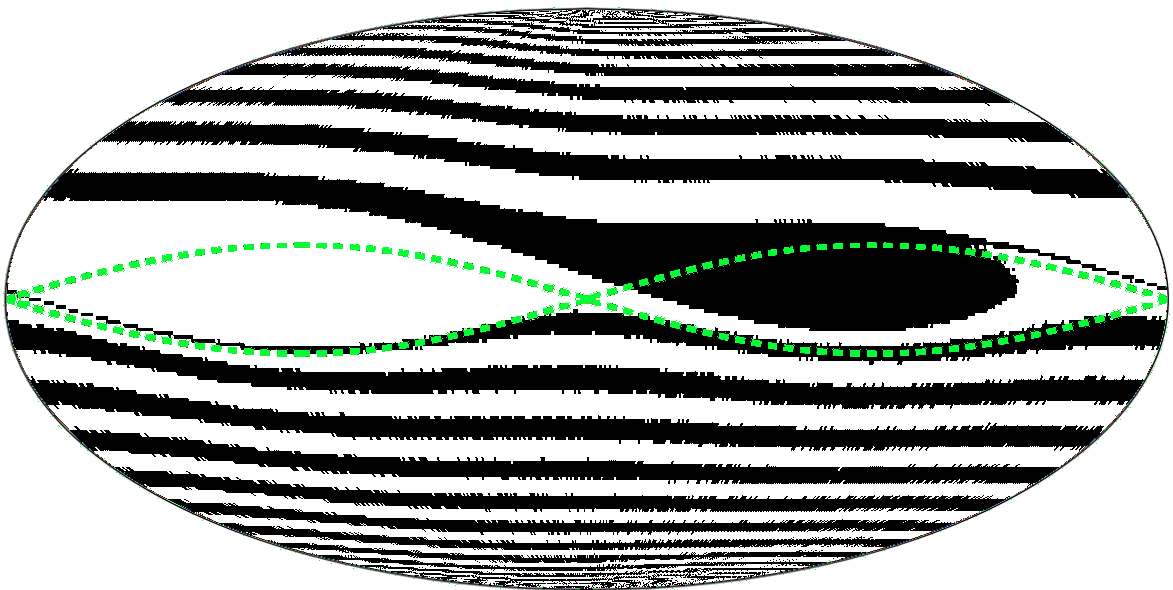} }}%
    \caption{{\footnotesize Relaxed configuration of the Bloch sphere as a function of initial magnetization for $D=10$, $\alpha=0.04$, $\omega=2.5\,\omega_c$ and temperature $\xi=5714$, where colors white/black imply relaxation into P/AP basins respectively. Imposing an exponential decay of a current pulse with fixed decay constant $\tau_I=0.01\,(n\mathrm{s}\cdot T)$ (Physical times are obtained upon dividing by $\mu_0H_K$), we vary the initial current intensity (a) $I=1.5I_c$ (b) $2.5I_c$ and (c) $3.5I_c$,  to show that the relaxation biasing effect observed numerically does not depend on the initial current intensity driving the system.}}
    \label{F7}%
\end{figure}

The model we have described further predicts that the final magnetization state only depends on the rate at which the current is decreased, not the initial value of the current that sustains the OPP orbits. We have confirmed this result by conducting numerical simulations for different initial currents with the same relaxation rate of the current. The results are shown in Fig.~\ref{F7} for initial currents of $I=1.5,2.5,3.5\,I_c$ and a relaxation time of $\tau_I=0.01\,(n\mathrm{s}\cdot T)$. We see that changing the initial current intensity has no effect on the biasing of the relaxation. Thus the orthogonal drift biasing has been shown to depend exclusively on the LLGS dynamics when the current is varied adiabatically.

\section{Conclusion}
In summary, we have demonstrated that an adiabatically decreasing current pulse leads to highly reliable precessional switching and explained this phenomena within the context of a macrospin model, identifying the time-scales that govern adiabatic current variations. These results can be tested on orthogonal spin-transfer torque devices as well as other types of spin-transfer torque oscillators. Our theory makes specific predictions for the switching probability as a function of the pulse decay time and temperature. The model also makes a strong prediction that the switching probability will be independent of the initial current that sustains the out-of-plane precessional orbit. Further for a slowly decaying current pulse the final magnetization state is also insensitive to the pulse shape and area, only depending on the pulse polarity.
\section*{Acknowledgments}

The authors would like to acknowledge J.-V. Kim, E. Vanden-Eijnden, K. Newhall and D.L. Stein for useful discussions. This research is based upon work supported by the Office of the Director of National Intelligence (ODNI), Intelligence Advanced Research Projects Activity (IARPA), through the U.S. Army Research Office under Contract No. W911NF-14-C-0089. It was also supported in part by National Science Foundation 
under Contract No. NSF-DMR-1309202. The content of the information does not necessarily
reflect the position or the policy of the Government, and no official endorsement should be inferred.

\bibliographystyle{unsrt}
\bibliography{mybib,Kentbib}

\begin{thebibliography}{10}

\bibitem{Ralph2008}
D.C. Ralph and M.D. Stiles.
\newblock Spin transfer torques.
\newblock {\em Journal of Magnetism and Magnetic Materials}, 320(7):1190,
  (2008).

\bibitem{BrataasKentOhno2012}
A.~Brataas, A.~D. Kent, and H.~Ohno.
\newblock Current-induced torques in magnetic materials.
\newblock {\em Nature Materials}, 11:372, 2012.

\bibitem{Silva2008}
T.J. Silva and W.H. Rippard.
\newblock Developments in nano-oscillators based upon spin-transfer
  point-contact devices.
\newblock {\em Journal of Magnetism and Magnetic Materials}, 320(7):1260,
  (2008).

\bibitem{KentWorledge2015}
Andrew~D. Kent and Daniel~C. Worledge.
\newblock A new spin on magnetic memories.
\newblock {\em Nature Nanotechnology}, 10(3):187--191, 2015.

\bibitem{Newhall2013}
K.~Newhall and E.~Vanden-Eijnden.
\newblock Averaged equation for energy diffusion on a graph reveals bifurcation
  diagram and thermally assisted reversal times in spin-torque driven
  nanomagnets.
\newblock {\em J. Appl. Phys.}, 113:184105, 2013.

\bibitem{Pinna2013}
D.~Pinna, A.~D. Kent, and D.~L. Stein.
\newblock Thermally assisted spin-transfer torque dynamics in energy space.
\newblock {\em Phys. Rev. B}, 88:104405, 2013.

\bibitem{Pinna2014}
D.~Pinna, D.~L. Stein, and A.~D. Kent.
\newblock Spin-torque oscillators with thermal noise: A constant energy orbit
  approach.
\newblock {\em Phys. Rev. B}, 90:174405, 2014.

\bibitem{Kent2004}
A.~D. Kent, B.~Ozyilmaz, and E.~del Barco.
\newblock Spin-transfer-induced precessional magnetization reversal.
\newblock {\em Applied Physics Letters}, 84(19):3897--3899, 2004.

\bibitem{Ye2015}
Li~Ye, Georg Wolf, Daniele Pinna, Gabriel~D. Chaves, and Andrew~D. Kent.
\newblock State diagram of an orthogonal spin transfer spin valve device.
\newblock {\em Journal of Applied Physics}, 117(19):193902, (2015).

\bibitem{Chaves2015}
Gabriel~D. Chaves-O'Flynn, Georg Wolf, Daniele Pinna, and Andrew~D. Kent.
\newblock Micromagnetic study of spin transfer switching with a spin
  polarization tilted out of the free layer plane.
\newblock {\em Journal of Applied Physics}, 117:17D705, (2015).

\bibitem{Karatsas}
I.~Karatsas and S.~Shreve.
\newblock {\em Brownian Motion and Stochastic Calculus}.
\newblock Springer-Verlag and New York, 2 edition, (1997).

\bibitem{Palacios}
Jos\'e~Luis Garc\'{\i}a-Palacios and Francisco~J. L\'azaro.
\newblock Langevin-dynamics study of the dynamical properties of small magnetic
  particles.
\newblock {\em Phys. Rev. B}, 58:14937--14958, Dec (1998).

\bibitem{LiZhang}
Z.~Li and S.~Zhang.
\newblock Thermally assisted magnetization reversal in the presence of a
  spin-transfer torque.
\newblock {\em Phys. Rev. B}, 69:134416, Apr (2004).

\bibitem{Apalkov}
D.~M. Apalkov and P.~B. Visscher.
\newblock Spin-torque switching: Fokker-planck rate calculation.
\newblock {\em Phys. Rev. B}, 72:180405, (2005).

\bibitem{PinnaIEEE}
D.~Pinna, D.~L. Stein, and A.~D. Kent.
\newblock Uniaxial thermally-assisted spin-transfer torque magnetization
  reversal in energy space.
\newblock {\em IEEE Trans. Mag.}, {\bf 49}(7):3144, (2013).

\bibitem{Molleweide}
J.P. Snyder.
\newblock {\em Flattening the Earth: Two Thousand Years of Map Projections}.
\newblock Chicago Press, (1993).

\end{thebibliography}
\newpage
\appendix
\section{Orthogonal Drift Dynamics on the Separatrix}
Here we derive the projections of the LLGS dynamics~(\ref{eq:Langevindynamics}) on the $\epsilon=0$ separatrix starting from the LLGS equation:
\be
\dot{\mathbf{m}}=\mathbf{m}\times\mathbf{h}_{\mathrm{eff}}-\alpha\mathbf{m}\times\left(\mathbf{m}\times\mathbf{h}_{\mathrm{eff}}\right)-\alpha I\mathbf{m}\times\left(\mathbf{m}\times\mathbf{\hat{n}}_p\right),
\ee
To do so we first rewrite the parametrization employed in Eqn.~(\ref{eq:SepParam}):
\bea
\bm{\gamma}^{1,2}(s)&=&\left(\pm\sqrt{\frac{D}{D+1}}\sin(s),\cos(s),\frac{1}{\sqrt{D+1}}\sin(s)\right)\\
\bm{\gamma}_{//}^{1,2}(s)&=&\left(\pm\sqrt{\frac{D}{D+1}}\cos(s),-\sin(s),\frac{1}{\sqrt{D+1}}\cos(s)\right)\\
\bm{\gamma}_{\bot}^{1,2}(s)&=&\frac{1}{\sqrt{D+1}}\left(1,0,\mp\sqrt{D}\right).
\eea 
The effective field for a biaxial macrospin is given by $\mathbf{h}_{\mathrm{eff}}=-\bm{\nabla}\epsilon=-\bm{\nabla}_{\mathbf{m}}\left(Dm_z^2-m_x^2\right)$ and the spin-polarization axis $\mathbf{\hat{n}}_p=(\cos\omega,0\sin\omega)$ is tilted by an angle $\omega$ with respect to the easy-axis. On the separatrix:
\bea
\mathbf{m}\times\mathbf{h}_{\mathrm{eff}}&=&-2\left(\frac{D}{\sqrt{D+1}}\sin(s)\cos(s),\mp\sqrt{D}\sin^2(s),\pm\sqrt{\frac{D}{D+1}}\sin(s)\cos(s)\right)\\
\mathbf{m}\times\mathbf{\hat{n}}_p&=&\left(\sin\omega\cos(s),\frac{\cos\omega}{\sqrt{D+1}}\left(1\mp\sqrt{D}\tan\omega\right),\cos\omega\cos(s)\right),
\eea 
where the portion of the separatrix bounding the $\mathrm{OOP}^+$ basin corresponds to $s\in[0,\pi]$. Employing the vector identities $\bm{\gamma}_{\bot}\cdot\left(\mathbf{m}\times\bm{A}\right)=\bm{A}\cdot\left(\bm{\gamma}_{\bot}\times\mathbf{m}\right)=\bm{A}\cdot\bm{\gamma}_{//}$ (conversely $\bm{\gamma}_{//}\times\mathbf{m}=-\bm{\gamma}_{\bot}$) where $\bm{A}$ is any vector, gives Eqn.~(\ref{eq:SepDynamics}) in the main text.

\section{Energy Relaxation Dynamics}

In this appendix we derive the timescale for magnetic relaxation onto a stable OOP limit cycle orbit with energy $\epsilon_0$ consistent with some driving current $I$. First we note that form of the energy equation is as follows~\cite{Pinna2014}:
\be
\dot{\epsilon}=-\alpha f_D\left(\epsilon,\tilde{I}\right)+g(\epsilon)\cdot\dot{W}.
\ee
For a macrospin precessing in the $\mathrm{OOP}^+$ basin this equation is ~\cite{Pinna2014}:
\bea
\label{eq:eevolve}
\partial_{\mathrm{t}}\epsilon(\gamma)&=&\frac{\pi\alpha}{\eta_0(\gamma)}\frac{D(D+1)}{[D(1-\gamma^2)+1]^{3/2}}\nonumber\\
&\times&\left\{\pm \tilde{I}(1-\gamma^2)-\frac{2}{\pi}\sqrt{D(1-\gamma^2)+1}\left[\eta_1(\gamma)-\frac{\gamma^2}{(D(1-\gamma^2)+1)}\eta_0(\gamma)\right]\right\}\nonumber\\
&+&h(\epsilon)\nonumber\\
&+&\sqrt{\frac{2\alpha}{\xi}\frac{D(D+1)}{D(1-\gamma^2)+1}\frac{1}{\eta_0(\gamma)}\left(\eta_1(\gamma)-\frac{\gamma^2}{D(1-\gamma^2)+1}\eta_0(\gamma)\right)}\cdot \dot{W_{\epsilon}}\\
h(\epsilon)&=&\frac{\alpha}{\xi}\frac{D(1-\gamma^2)+1}{1-\gamma^2}\left[1-\left(\frac{D(1-\gamma^2)+2}{D(1-\gamma^2)+1}\right)\frac{\mathrm{E}[1-\gamma^2]}{\mathrm{K}[1-\gamma^2]}+\frac{1}{\gamma^2(2-\gamma^2)}\left(\frac{\mathrm{E}[1-\gamma^2]}{\mathrm{K}[1-\gamma^2]}\right)^2\right]\nonumber\\
&+&\frac{\alpha}{\xi}\frac{D(1+\gamma^2)+1}{D(1-\gamma^2)+1},
\eea
where $\eta_0(\gamma)=\mathrm{K}[1-\gamma^2]$ and $\eta_1(\gamma)=\mathrm{E}[1-\gamma^2]$ are expressed in terms of complete elliptic integrals of the first and second kind, $\gamma(\epsilon)=\epsilon(D+1)/\left[D(1+\epsilon)\right]$ depends on the energy $\epsilon$ and $h(\epsilon)$ is a drift-diffusion correction term.

Upon linearizing~(\ref{eq:eevolve}) around an $\epsilon_0$ fixed point ($\epsilon\to\epsilon_0+\delta\epsilon$), the deterministic portion of the dynamics governing perturbations $\delta\epsilon$ will be to first order:
\be
\dot{\delta\epsilon}=-\alpha \left[\partial_{\epsilon}f_D(\epsilon,\tilde{I})\right]_{\epsilon=\epsilon_0(I)}\delta\epsilon,
\ee
whose solution is:
\bea
\delta\epsilon(t)&=&\delta\epsilon_0\exp\left[-\frac{t}{\tau_I}\right]\\
\tau_I&=&\left[\partial_{\epsilon}f_D(\epsilon,\tilde{I})\right]_{\epsilon=\epsilon_0(I)}^{-1}.
\eea
\end{document}